\documentclass[%
reprint,
superscriptaddress,
a4paper,
showpacs,preprintnumbers,
amsmath,amssymb,
aps,
prstab,
floatfix,
]{revtex4-1}

\usepackage{graphicx}
\usepackage[caption=false]{subfig}
\usepackage{float}
\usepackage{hyperref}
\hypersetup{colorlinks=true,citecolor=blue}

\begin{document}
\title{Low emittance muon beam in the 2 to 40~GeV energy range for muon and neutrino experiments}
\author{O.R. Blanco-Garc\'ia}
\email{oblancog@lnf.infn.it}
\affiliation{INFN-LNF, Via E.~Fermi 54 (gi\`a 40), 00044 Frascati, Rome, Italy}
\date{\today}

\begin{abstract}
  I present a scheme to obtain a 2 to 40~GeV low emittance muon beam, not requiring cooling and within today's technological resources, to be used for early commissioning of muon accelerator projects, or alternatively dedicated muon and neutrino parameter measurements.\par
  In particular, a muon rate of $5\times10^4$~$\mu$/s in a normalized transverse emittance of $5$~$\pi$~$\mu$m at 22~GeV, and energy spread of 1~GeV obtained from $O(10^{11})$~e$^+$/s on target at 44~GeV. This emittance is below the expected results of advanced emittance cooling techniques for muons produced from protons-on-target, and represents an alternative for the duration of complete muon cooling studies.\par
  The scheme has beam designed to adjust the muon beam energy in the GeV energy range to the needs for precise parameter measurements of muons and neutrinos. Although the rate is small compared to other muon sources, it does not seem to represent a big limitation for its usage.\par
  Furthermore, the muon rate could be in principle increased proportionally to the availability of higher positron rates, already foreseen for future collider projects.\par

\end{abstract}
\maketitle
\section{Introduction}
There is an increasing interest in the accelerator community to study a high luminosity and high energy muon--muon collider that could continue the exploration of the particle physics energy frontier using leptons~\cite{muoncolliders}, while studies of a muon source based on protons-on-target~(PoT) collisions~\cite{map} show that it could achieve, in principle, an instantaneous luminosity above $10^{34}$~cm$^{-2}$~s$^{-1}$ in the TeV energy range from $O(10^{12})$ muons per bunch inside an emittance foreseen to be reduced with advanced muon emittance cooling concepts.\par
The results of the initial test limited to conclude on the muon beam transverse emittance have been recently released~\cite{mice} giving positive indications of emittance cooling, while, the total combination of transverse and longitudinal emittance cooling remains yet untested.\par
Alternatively, a low emittance muon beam can be generated from positrons on target. These studies are carried out by the LEMMA~(Low EMittance Muon Accelerator)~\cite{NIM,eplusringopt,MaricaIPAC2019,alesini2019positron} team at INFN/Italy, who is interested in creating a muon bunch population in the order of $10^9$ particles in an extremely small normalized transverse emittance $\epsilon_n$ of $0.04\;\pi\;\mu$m at 22.5~GeV and longitudinal emittance of about $3\times1$~$\pi$~mm~GeV, to produce luminosities in the order of $10^{34}$~cm$^{-2}$~s$^{-1}$.\par
Recent results on the LEMMA muon source~\cite{blancoprab2020} have been able to show a ten fold increase in the muon production efficiency, reaching $0.5\times10^{-6}$~$\mu$~pairs per impinging positron~($\mu/e^+$), in a reduced transverse normalized emittance of $5\;\pi\;\mu$m at 22~GeV and longitudinal emittance of $3\;\pi\;$mm~GeV, while at the same time reducing the power deposition per target by more than a factor ten. Although the transverse emittance attained is still larger than the LEMMA goal, it already offers the possibility of a muon source with an emittance below the one foreseen from protons after cooling. Therefore, the studies of the LEMMA muon source provide an alternative to cooling available with today's technology.\par
Muon sources have been used or suggested for studies on muon and neutrino parameters~\cite{boscolo1808} along varying energy and intensity. For example, accelerator facilities like nuSTORM~(Neutrinos from STORED Muons)~\cite{nustorm,nustormatCERN,mudecay}, nuMAX~(Neutrinos from a Muon Accelerator Complex)~\cite{numax}, the Muon $g-2$ experiment~\cite{gminus2} and MUonE~\cite{muone} have been designed in the GeV energy range considering larger muon emittances or projecting in the future the usage of a small emittance given that results from cooling are favorable.\par
Instead, muon facilities could benefit from having today a low emittance muon beam to increase the precision of their measurements, reduce the experiment complexity or even consider and earlier commissioning at low intensity.\par
For example, we could expect to have an instantaneous luminosity of $10^{33}\;$cm$^{-2}$~s$^{-1}$ from the collision of a muon beam with the electrons in a high density fixed target~(similar to the MUonE experiment). Let's assume a muon flux $\Phi$ of $5\times10^4$~$\mu$/s, impinging on a target $l$ 40~m long, density $\rho_m$ of 20~g cm$^{-3}$, from which we get the luminosity~\cite{widermann}
\begin{equation}
  L = \Phi\rho l = 1.2\times10^{33}\text{ cm}^{-2}\text{s}^{-1},
\end{equation}
where we have use the Avogadro Number $N_a$ (approximately $6\times10^{23}$ nuclei/mol) and assumed a factor 1/2 coming from the ratio Z/A of the material atomic number $Z$ to the  mass number $A$, in order to calculate $\rho=\rho_m\frac{Z}{A}N_a$.\par
Note that for a muon beam with energy varying from 2 to 40~GeV on a fixed target collision, the center of mass energy is in the range of 115 to 228~MeV.\par
Although some parameters might seem excessive at first glance, I would like to highlight the benefit of a low emittance muon beam.\par
First, it allows to have a high muon flux density reducing the requirements on the target transverse dimensions. Supposing that the accelerator optics twiss beta function at the target is $\beta=10$~m, we can approximate the beam dimension to $\sigma=\sqrt{\beta\epsilon_n/\gamma_r}=0.5$~mm and divergence $\sigma'=\sqrt{\epsilon_n/(\beta\gamma_r)}=0.05$~mrad for a relativistic gamma  $\gamma_r=200$~(an ultra-relativistic muon at approximately 22~GeV). Therefore, the 40~m long target can be contained in a volume of 40~m $\times$ 5~mm $\times$ 5~mm.\par
The lifetime of the muon beam at 22~GeV is as long as 0.4~ms which gives the possibility to consider a ring to recirculate the muon beam through a shorter target, as in~\cite{oscarmuacc,PhysRevAccelBeams.23.051001}, possibly rising the luminosity by a factor 10 to 100.\par
Furthermore, the muon rate of $5\times10^4$~$\mu$/s could be obtained from a positron rate of $10^{11}$~$e^+$/s that is within the reach of several laboratories around the world, and higher muon rates could be achieved with dedicated efforts to design a custom  positron source.\par
With respect to neutrino experiments, previous measurements of interesting neutrino cross sections~\cite{nucross} report values well above $10^{-39}$~cm$^2$ at 1~GeV. Estimating 1 year as $2\times10^7$~s, in 5 years of run we should be able to detect a large number of the 100 muon to neutrino decay events in the GeV energy range. In spite of the small number of events, the scheme provides two main advantages: the high muon flux density and the energy tunability.\par
The muon flux is as high as $2\times10^7$~$\mu$~s$^{-1}$~cm$^{-2}$ from $5\times10^4$ muons per second on $0.5$~mm$\times0.5$~mm. A similar number of neutrino events could occur when muons are left to decay in a ring. In contrast, atmospheric neutrinos in the GeV scale would have a neutrino flux of 1 to 0.001 s$^{-1}$~cm$^{-2}$, see Fig.~4.1 of~\cite{flux}, assuming a solid angle of 1~sr and 1000~MeV beam energy spread.\par
Furthermore, tuning the muon beam energy in the range of 2 to 40~GeV gives the possibility to create neutrino events in the desired energy range.\par
The presented case for neutrino statistics is marginal, but, higher neutrino rates could be achieved by improvements in the positron source. A factor 100 higher positron rate, of $0.6\times10^{13}$~e$^+$/s, has been reported at SLAC for the Standford Linear Collider~SLC~\cite{possource}, and rates of $O(10^{14})$~e$^+$/s are currently foreseen for future collider projects.\par
In addition, higher neutrino rates could be achieved by increasing a factor 2 or 4 the muon production efficiency, and by optimizations of the muon decay ring for better performance with low emittance beams.\par
In general, many applications could derive from a facility dedicated to a low emittance source constructed while cooling experiments are conclusive, e.g. muon tomography at varying the beam energy in the GeV range~\cite{muontomography,lechmann} or neutrino tomography for the earth nucleus studies~\cite{nutomography}.\par
\section{The production scheme}
I would like to present, as input for further discussion, the production scheme of a low emittance muon beam in the energy range of 2 to 40~GeV as a source of muons and neutrinos.\par
Muons are produced from positron--electron annihilation of a 44~GeV~$e^+$ beam in multiple fixed targets, producing a small normalized muon beam emittance of 5~$\pi$~$\mu$m at 22~GeV without the need of cooling, therefore, the required technology is available today.\par
The muon generation sequence is shown in Eq.~(\ref{eq:seq}), it starts with low energy electrons that are accelerated to a fixed target where $e^+e^-$ pairs are produced. Positrons are captured and accelerated to be injected in a small damping ring which reduces the beam emittance. The positron bunch is extracted, accelerated to 44~GeV and passed through a series of targets connected by a transport line that preserves the emittance of the newly created muon beam and mitigates the degradation of the positron beam. One of the two muon species is selected to be transported for experiments.\par
\begin{figure*}[h]
  \includegraphics[width=0.98\textwidth,angle=0]{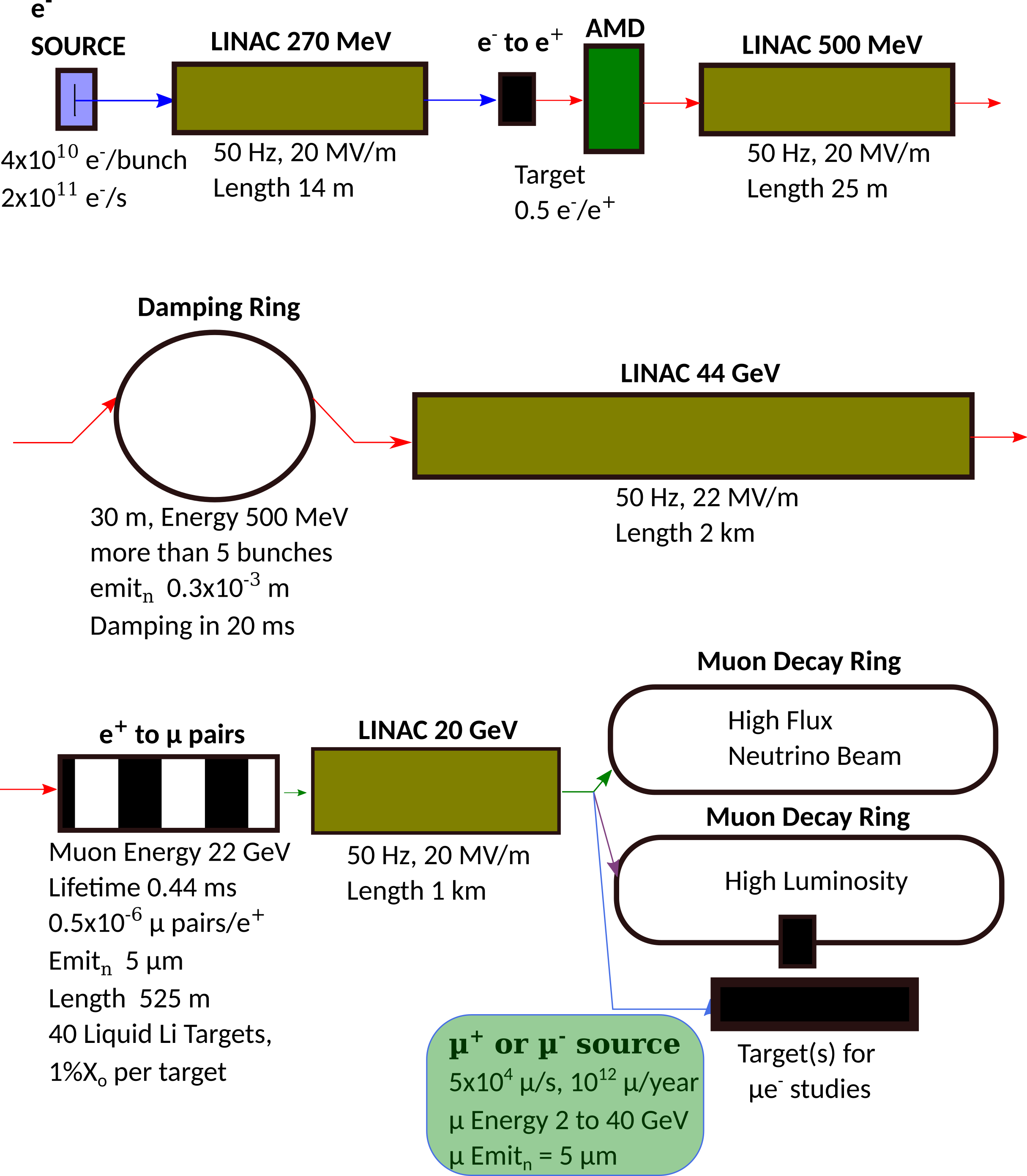}
  \caption{Low emittance muon beam production scheme for muon and neutrino studies.}\label{f:muscheme}
\end{figure*}

Details of the scheme are shown in Fig.~\ref{f:muscheme}, where the main parameters are display on each stage.\par
We initially follow a positron production scheme similar to that used for DA$\Phi$NE~\cite{dafne} at LNF/INFN in Italy, or that of VEPP-5~\cite{vepp5} at BINP in Russia, or KEKB~\cite{kekb} at KEK in Japan. We produce bunches of O($10^{10}$) electrons and accelerate them in a LINAC working at 50~Hz to few hundred MeVs (in DA$\Phi$NE the electron bunch reaches 190~MeV while for VEPP-5 it is accelerated to 270~MeV). This configuration provides a source of O($10^{11})$~e$^-$/s.\par
The electron beam interacts with a Tungsten, or Tantalum target to produce approximately 0.1 to 0.5~positrons per impinging electron after the capture and focusing section~(most likely with an Adiabatic Matching Device, AMD). We assume from here on that the electron intensity and energy is enough to provide $10^{11}$~$e^+$/s.\par
The $10^{11}$~$e^+$/s are accelerated to 0.5~GeV in order to damp the beam emittance in a small damping ring~(D.R.), e.g. the 30~m long D.R. at the DA$\Phi$NE complex. In principle, a transverse normalized emittance of $0.3$~$\pi$~mm  would suffice to produce a small emittance muon beam (if we consider $\gamma_r=10^3$, the geometrical emittance at the D.R. is in the micrometer level).\par
The low emittance positron beam is accelerated in a 2~km LINAC to 44~GeV, reducing the geometrical emittance to the nanometer scale ($\gamma_r\approx0.9\times10^5$). Having positrons above the muon production energy threshold, at 43.7~GeV, we pass the positron bunch through 20 to 40 Liquid Lithium targets of 1\% of a radiation length~$X_0$ each, in order to dissipate in multiple locations the energy deposited by the beam while increasing the muon population by a factor 10, see the transport line design in~\cite{blancoprab2020}.\par
\begin{widetext}
  \begin{equation}
  e^- \xrightarrow[\text{to 0.3 GeV}]{\text{LINAC}} Target \rightarrow e^+ \xrightarrow[\text{to 0.5GeV}]{\text{LINAC}} \xrightarrow[\text{$e^+$ Damping}]{\text{Ring}} \text{Low emit. }e^+ \xrightarrow[\text{to 44~GeV}]{\text{LINAC}}Multiple\;Targets \rightarrow \text{Low emit. }\mu\text{ beam}\label{eq:seq}
  \end{equation}
\end{widetext}
At the end of the line we obtain $5\times10^4$~$\mu$/s, splitted in a train of 50 bunches of $10^3$~$\mu$ pairs at 22~GeV with a small geometrical transverse emittance of 25~$\pi$~nm, a small bunch length given by the positron damping ring and energy spread of $\pm$5\%.\par
\section{Muon beam usage}
It is possible to select the $\mu^+$ or the $\mu^-$ beam to be transported/or deflected towards next stages. The muon beam could collide with a target to study $\mu e^-$ interactions with luminosity of $10^{33}$~cm$^{-2}$~s$^{-1}$, or injected into a recirculation ring intercepted by the target to increase the luminosity by a factor 10 to 100.\par
Alternatively one could think to remove the target and let the beam decay to provide a high flux of muonic neutrinos for tomography or neutrino parameter studies. It could be possible to install two experiments, each one using one of the muon species.\par
For completeness, Table~\ref{t:listmat} shows a list of target materials and densities in the range of 20~g~cm$^{-3}$.\par
\begin{table}[htb]
  \centering
  \caption{List of materials with density near 20~g/cm$^3$.}\label{t:listmat}
    \begin{tabular}{lll}\hline
      Material & Symbol & Density \\
               &        & (g/cm$^3$)\\\hline
      Molybdenum & Mo& 10.28\\
      Silver & Ag & 10.49\\
      Lead & Pb & 11.34\\
      Thorium & Th & 11.7\\
      Rhodium & Rh & 12.41\\
      Mercury & Hg & 13.53\\
      Tantalum & Ta & 16.69\\
      Uranium & U & 19.1\\
      Tungsten & W & 19.25\\
      Gold & Au & 19.30\\
      Plutonium & Pu &19.85\\
      Rhenium & Re & 21.02\\
      Platinum & Pt & 21.45\\
      Iridium & Ir&22.56\\
      Osmium & Os & 22.59
    \end{tabular}
\end{table}

\section{Conclusion}
I have presented a scheme to produce a low emittance muon beam not requiring cooling and within today's technology resources.\par
The attained rate and normalized emittance of $5\times10^4$~$\mu$/s and $5$~$\pi$~$\mu$m, respectively, could be of use for an early commissioning of current muon beam projects in the 2 to 40~GeV energy range, or to design newly projects related to muon or neutrino parameter measurements that could profit from the reduced emittance.\par
The reduced muon rate obtained with positrons on target, when compared to protons-on-target, seems not to be an obstacle to achieve relevant luminosity for $\mu e^-$ collision studies and moderate neutrino rates, and further improvements in positron sources for new collider projects would be of benefit to increase the muon rate by one or maybe two orders of magnitude.\par
\section{Acknowledgments}
This work has been financially supported by the Istituto Nazionale di Fisica Nucleare~(INFN), Italy, Commissione Scientifica Nazionale~5, Ricerca Tecnologica -- Bando n.~20069.\par

\bibliographystyle{unsrt}

\end{document}